\documentclass[aps,prl,twocolumn,showpacs]{revtex4}

\usepackage{graphicx}

\usepackage{amsmath} \newcommand{\EqLabel}[1]{\label{#1}}

\begin{document}
 
\title{Green's function of a dressed particle}

\author{Mona Berciu}

\affiliation{Department of Physics and Astronomy, University of
  British Columbia, Vancouver, BC V6T 1Z1, Canada }

\date{\today}
 
\begin{abstract}
We present a new, highly efficient yet accurate approximation for the
Green's functions of dressed particles, using the Holstein polaron as
an example. Instead of summing a subclass of diagrams ({\em e.g.} the
non-crossed ones, in the self-consistent Born approximation
(SCBA)), we {\em sum all the diagrams}, but with each diagram averaged
over its free propagators' momenta. The resulting Green's function
satisfies exactly the first six spectral weight sum rules. All higher
sum rules are satisfied with great accuracy, becoming asymptotically
exact for coupling both much larger and much smaller than the free
particle bandwidth. Possible generalizations to other models are also 
discussed.
\end{abstract}


\maketitle

One of the most fundamental problems in both high-energy and condensed
matter physics is to understand what happens when a particle couples
to an environment, in particular what are the properties of the
resulting object, consisting of the bare particle dressed by a cloud
of excitations. This type of problem arises again and again as
couplings to new kinds of environments are studied.

The most desirable quantity to know is the Green's function
$G(\vec{k},\omega)$ of the dressed particle -- its poles mark the
eigenspectrum, while the associated residues contain information on
the eigenfunctions. Moreover, the spectral weight $A(\vec{k},\omega) =
-{1\over \pi} \mbox{Im} G(\vec{k},\omega)$ can be directly measured
experimentally using Angle-Resolved Photoemission
Spectroscopy~\cite{Andrea}. Recently, such work has reignited a debate
on whether the carriers in high-T$_{\rm c}$ cuprates are polarons,
that is, electrons dressed by phonons~\cite{hightc}.

$G(\vec{k},\omega)$ is the sum of an infinite number of diagrams
corresponding to an expansion to all orders in the coupling
strength~\cite{Mahan}. Diagrammatic Monte Carlo
(DMC) can perform the {\em numerical} summation of all
diagrams~\cite{QMC}.  Other ways to find $G(\vec{k},\omega)$ are from
exact diagonalizations (ED) of small systems, variational methods,
Density Matrix Renormalization Group in one-dimension,
etc~\cite{var,EDvar,DMFT,Cat}. However, these methods 
require considerable computational resources, are time consuming, and
often limit themselves to finding only the low-energy properties, such
as the ground-state energy.

\begin{figure}[b]
\includegraphics[width=0.99\columnwidth]{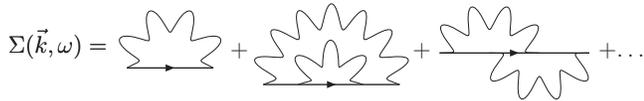}
\caption{Diagrammatics for $\Sigma$, where $G = G_0 + G_0 \Sigma G$.}
\label{fig1}
\end{figure}
To our knowledge, there are only two easy-to-estimate approximations
for $G(\vec{k},\omega)$. One is the the SCBA which consists in summing
only the non-crossed
diagrams. Because the percentage of diagrams kept decreases fast with
increasing order, SCBA fails badly at strong couplings. The other
approximation, obtained from a modified Lang-Firsov (MLF) approach
\cite{Alexandr}, is exact both for zero coupling and for zero
bandwidth, however results for finite bandwidth/coupling are
rather poor (see below).

In this Letter we find a new approximation for $G(\vec{k},\omega)$,
which is as easy to estimate as SCBA and MLF, but is {\em highly
accurate over most of the parameter space}. We validate this both by
comparison against numerical results, and by investigating its sum
rules. Most of the discussion here is limited to the Holstein model,
for which many numerical results are available. Possible
generalizations for other models are briefly discussed at the end.

Consider the Holstein Hamiltonian:
\begin{equation}
\EqLabel{1} {\cal H} \!= \!\!\sum_{\vec{k}}^{}( \epsilon_{\vec{k}}
c^{\dagger}_{\vec{k}} c_{\vec{k}} + \omega_E b^{\dagger}_{\vec{k}}
b_{\vec{k}})+ \frac{g}{\sqrt{N}}
\sum_{\vec{k},\vec{q}}^{}c^{\dagger}_{\vec{k}-\vec{q}} c_{\vec{k}}
(b^{\dagger}_{\vec{q}} + b_{-\vec{q}} )
\end{equation}
which contains the kinetic energy of the free particle (electron), a
branch of Einstein bosons (optical phonons) and the linear coupling
between particle and  bosons.  When needed, we use
$\epsilon_{\vec{k}} = - 2t\sum_{\alpha=1}^{d} 
\cos(k_\alpha a)$, for
nearest-neighbor hopping on a d-dimensional cubic lattice with $N$ sites
and lattice constant $a$. The spin of the particle and whether it is a
boson or a fermion is irrelevant. Sums over momenta are over the first
Brillouin zone (BZ), $-{\pi\over a}< k_\alpha \le {\pi\over a}$, $\alpha=1,d$.

The one-particle Green's function  of a $M$-particle
system is $G(\vec{k},\!\tau) = -i \langle \Phi_M| T
[c_{\vec{k}}(\!\tau) c_{\vec{k}}^{\dagger}(0) ]|\Phi_M\rangle$, with
$|\Phi_M\rangle$ the $M$-particle ground-state, $T$ the time ordering
operator, and $c_{\vec{k}}(\tau)=\exp(i {\cal H} \tau) c_{\vec{k}}
\exp(-i{\cal H} \tau)$ (we set $\hbar=1$)~\cite{Mahan}. The spectrum of a single
dressed particle (polaron) is obtained for $M=0$. The ground state
(GS) in the absence of particles is the vacuum $|\Phi_0\rangle
=|0\rangle$, and since ${\cal H}|0\rangle =0$, the polaron's Green's
function simplifies to:
\begin{equation}
\EqLabel{2} G(\vec{k}, \tau) = -i \Theta(\tau) \langle 0 | c_{\vec{k}}
 e^{-i {\cal H} \tau} c_{\vec{k}}^{\dagger} |0\rangle
\end{equation}
where $\Theta(\tau)$ is the Heaviside function. From Eqs. (\ref{1})
and (\ref{2}), we derive the equation of motion:
$$ i {d \over d\tau} G(\vec{k},\tau) = \delta(\tau) +
\epsilon_{\vec{k}} G(\vec{k},\tau) + {g \over \sqrt{N}}
\sum_{\vec{q}_1}^{} F_1(\vec{k}, \vec{q}_1,\tau)
$$
where $ F_1(\vec{k}, \vec{q}_1, \tau) = -i \Theta(\tau) \langle 0|
c_{\vec{k}} \exp({-i {\cal H} \tau})
c^{\dagger}_{\vec{k}-\vec{q}_1}b^{\dagger}_{\vec{q}_1}|0\rangle $.
Continuing in this vein, we generate an infinite hierarchy of coupled
equations of motion for the Green's functions $F_n(\vec{k}, \vec{q}_1,
\dots, \vec{q}_n, \!\tau)\! = \!- i \Theta(\tau) \langle 0|
c_{\vec{k}} e^{-i {\cal H} \tau}\!
c^{\dagger}_{\vec{k}-\vec{q}_T}b^{\dagger}_{\vec{q}_1}\cdots
b^{\dagger}_{\vec{q}_n}|0\rangle$, with
$\vec{q}_T=\sum_{i=1}^{n}\vec{q}_i$ and
$G(\vec{k},\tau)=F_0(\vec{k},\tau)$.  In the frequency domain, these
equations of motion become:
\begin{equation}
\EqLabel{5} G(\vec{k},\omega) = G_0(\vec{k},\omega) [1+ {g\over
\sqrt{N}}\sum_{\vec{q}_1}^{} F_1(\vec{k},\vec{q}_1,\omega) ]
\end{equation}
and for any $n\ge 1$,
\begin{eqnarray}
\nonumber F_n(\vec{k},\vec{q}_1,\dots,\vec{q}_n,\omega) = {g\over
  \sqrt{N}} G_0 ( \vec{k}-\vec{q}_T,\omega - n \omega_E)
  & &\\ \nonumber \times[\sum_{i=1}^{n}
  F_{n-1}(\vec{k},\vec{q}_1,\dots,\vec{q}_{i-1},\vec{q}_{i+1},\dots,\vec{q}_n,\omega)
   \hspace{10mm}& &\\ \EqLabel{7} + \sum_{\vec{q}_{n+1}}^{}
  F_{n+1}(\vec{k},\vec{q}_1,\dots,\vec{q}_n,\vec{q}_{n+1},\omega) ]
\end{eqnarray}
where $G_0(\vec{k},\omega) = [\omega - \epsilon_{\vec{k}} +i
\eta]^{-1}$ is the non-interacting one-particle Green's function. This
system generates the expected diagrammatic expansion for the
self-energy $\Sigma(\vec{k},\omega)= G_0^{-1}(\vec{k},\omega) -
G^{-1}(\vec{k},\omega)$, shown in Fig. \ref{fig1}.

Let $ f_n(\vec{k},\omega) = N^{-n}\sum_{\vec{q}_1,\dots,\vec{q}_n}^{}
F_n(\vec{k}, \vec{q}_1,\dots,\vec{q}_n,\omega) $. In terms of these,
Eq. (\ref{5}) becomes $G(\vec{k},\omega) = G_0(\vec{k},\omega)[1+g
\sqrt{N} f_1(\vec{k},\omega) ]$. The equations for
$f_n(\vec{k},\omega)$, $n \ge 1$, are obtained by summing
Eqs. (\ref{7}) over all phonon momenta. Of the two terms on the
right-hand side, the first one can be expressed in terms of
$f_{n-1}(\vec{k},\omega)$ exactly, but the second one requires an
approximation. We replace:
\begin{eqnarray}
\nonumber
&&\!\!\!\!\!\!\sum_{\vec{q}_1,\dots,\vec{q}_{n+1}}\!\!\!\!\!\!
G_0(\vec{k}-\vec{q}_T,\omega-n\omega_E)
F_{n+1}(\vec{k},\vec{q}_1,\dots,\vec{q}_{n+1},\omega) \\ &&
\EqLabel{10a} \hspace{20mm} \approx N^{n+1} {\bar
g_0}(\omega-n\omega_E) f_{n+1}(\vec{k},\omega)
\end{eqnarray}
where
\begin{equation}
  \EqLabel{11} {\bar g_0}(\omega) = {1\over N} \sum_{\vec{k}}^{}
  G_0(\vec{k},\omega)
\end{equation}
The justification is that $\vec{q}_T= \sum_{i=1}^{n}\vec{q}_i$ takes,
with equal probability, any value in the first Brillouin
zone. Replacing $G_0(\vec{k}-\vec{q}_T, \omega - n\omega_E)
\rightarrow \langle G_0(\vec{k}-\vec{q}_T, \omega -
n\omega_E)\rangle_{\vec{q}_T} = {\bar g_0}(\omega-n\omega_E)$ allows
us to also write this term as a function of $f_{n+1}(k,\omega)$
only. We discuss below the meaning of this momentum average (MA) in
terms of diagrams; however, note that for hopping $t= 0$ this MA
approximation becomes exact, because for $t=0$ all Green's functions
are independent of momenta. This suggests that MA should be valid at
least in the strong-coupling regime $t/g \ll 1$. As we show later, its
validity range is in fact much wider.

With this approximation, Eqs. (\ref{7}) become $ f_n(\vec{k},\omega)\!
= \!{\bar g_0}(\omega-n\omega_E)\!\!\left[\frac{ng}{ \sqrt{N}}
f_{n-1}(\vec{k},\omega)\! + \!g \sqrt{N}
f_{n+1}(\vec{k},\omega)\right] $. This recursive chain has a
continued-fraction solution. The resulting Green's function can be
cast in the usual form $G_{MA}(\vec{k},\omega) = [ \omega -
\epsilon_{\vec{k}} - \Sigma_{MA}(\omega)+i \eta]^{-1}$, where:
\begin{equation}
\EqLabel{12} \Sigma_{MA}(\omega) = \cfrac{g^2 {\bar
g_0}(\omega-\omega_E)}{\Large 1 - \cfrac{2g^2 {\bar
g_0}(\omega-\omega_E) {\bar g_0}(\omega-2\omega_E)}{\large 1-
{\cfrac{3 g^2 {\bar g_0}(\omega-2\omega_E) {\bar
g_0}(\omega-3\omega_E)}{1-\cdots}}}}
\end{equation}
As pointed already out, if $t=0$, in which case ${\bar g_0}(\omega) = (\omega + i
\eta)^{-1}$, this expression is exact. Indeed, one can show
\cite{next} that it equals the expected Lang-Firsov
result~\cite{Mahan} [$\lambda=(g/\omega_E)^2$]:
\begin{equation}
\EqLabel{12b} G(\omega) = e^{-\lambda}
\sum_{n=0}^{\infty}\frac{\lambda^{n}}{n!}\frac{1}{\omega + \lambda
\omega_E - n \omega_E +i \eta}
\end{equation}
This mapping was used before in a Dynamical Mean Field Theory (DMFT) study of this
problem, which also produces an approximation for the Green's
function~\cite{DMFT}. In fact,  Eq. (\ref{12}) looks similar to
$\Sigma_{DMFT}(\omega)$; however, our $\bar{g}_0(\omega)$ is {\em not}
a solution of the self-consistent DMFT equations (except at $t=0$ and
$g=0$, where both methods are exact). At finite $g/t$ the two
self-energies are different. Moreover, because of the limit $d\rightarrow
\infty$, in DMFT $G$ itself (not only $\Sigma$) is independent of
$\vec{k}$. Finally, the DMFT evaluation requires self-consistent
iterations, and is therefore much more involved than that of the MA,
SCBA and MLF. For these reasons, we do not consider the results
of the DMFT in the following.

Interestingly, SCBA also depends on ${\bar g_0}(\omega)$. Since
$\Sigma_{SCBA}(\omega)={g^2\over N} \sum_{q}^{} G_{SCBA}(k-q,\omega
-\omega_E)$, we have:
\begin{eqnarray}
 \Sigma_{SCBA}(\omega) = g^2{\bar g_0}\left(\omega - \omega_E -
\Sigma_{SCBA}\left(\omega-\omega_E \right) \right)\hspace{5mm}
&&\nonumber\\ = g^2 {\bar g_0}\left(\omega \!- \!\omega_E \!-g^2{\bar
g_0}\left(\omega\!-\!2\omega_E-\! g^2{\bar g_0}(\omega\! -\! 3
\omega_E\! - \dots )\right)\right) \nonumber &&
\end{eqnarray}
On the other hand, for the Holstein model, the MLF expression is
reminiscent of Eq. (\ref{12b}) \cite{Alexandr,Korni}:
$$ G_{MLF}(\vec{k},\omega) = e^{-\lambda}
\sum_{n=0}^{\infty}\frac{\lambda^{n}}{n!}\frac{1}{\omega
-e^{-\lambda}\epsilon_{\vec{k}} +\lambda \omega_E - n \omega_E +i
\eta}
$$

To understand the diagrammatic meaning of the MA approximation, we expand
Eq. (\ref{12}) in powers of $g^2$:
\begin{eqnarray}
\nonumber \Sigma_{MA}(\omega)= g^2 {\bar g_0}(\omega-\omega_E) + g^4 2
{\bar g_0}^2(\omega-\omega_E) {\bar g_0}(\omega-2\omega_E)&& \\
\nonumber + g^6 \left[ 4{\bar g_0}^3(\omega-\omega_E) {\bar
g_0}^2(\omega-2\omega_E) + 6 {\bar
g_0}^2(\omega-\omega_E)\right.\hspace{8mm} && \\\label{sigma}
\left. \times{\bar g_0}^2(\omega-2\omega_E) {\bar
g_0}(\omega-3\omega_E)\right] +{\cal O}(g^8)\hspace{3mm} &&
\end{eqnarray}
showing one contribution of order $g^2$ (which is the correct Born
expression), 2 of order $g^4$, 10 of order $g^6$, etc. One can verify
that this generates the correct total number of diagrams in all
orders. The difference is that in all MA diagrams, each
$G_0(\vec{p},\Omega) $ free propagator is replaced by a momentum
averaged ${\bar g_0}(\Omega)$ function.  For example, the exact
$2^{nd}$ order contribution (see Fig. \ref{fig1}):
\begin{eqnarray}
\nonumber {g^4\over N^2} \sum_{\vec{q}_1,\vec{q}_2}^{}
G_0(\vec{k}-\vec{q}_1,\omega-\omega_E)
G_0(\vec{k}-\vec{q}_1-\vec{q}_2,\omega-2\omega_E) && \\
\label{ex} \times\left[ G_0(\vec{k}-\vec{q}_1,\omega-\omega_E) +
  G_0(\vec{k}-\vec{q}_2,\omega-\omega_E)\right]\hspace{5mm} &&
\end{eqnarray}
is replaced within the MA approximation by
\begin{equation}
\EqLabel{ap} 2g^4 \!\!\left(\!\!{1\over N}
  \!\sum_{\vec{q}_1}^{}G_0(\vec{q}_1,\omega\!-\!\omega_E) \!\!
  \right)^{\!\!2}\!\!\!\left(\!\!{1\over N}\!
  \sum_{\vec{q}_2}^{}G_0(\vec{q}_2,\omega\!-\!2\omega_E)\!\! \right)
\end{equation}
All higher orders are obtained similarly. Let us see why this is
indeed a good approximation if $t \ll g$. If $t=0$, the two
expressions are equal. Higher order powers of $t$ come from expanding
each $G_0(\vec{k},\omega) = G_0(\omega) + \epsilon_{\vec{k}}
G_0^2(\omega) + \dots$, where $G_0(\omega)=(\omega+i\eta)^{-1}$. All
odd-order powers are zero since $\sum_{\vec{k}}^{}
\epsilon_{\vec{k}}^{2n+1} =0$. Consider ${\cal O}(t^2)$ terms in
Eq. (\ref{ex}): these come either from expanding one of the $G_0$ to
${\cal O}(t^2)$, in which case they equal their counterparts in
Eq. (\ref{ap}); or they come from ${\cal O}(t)$ contributions from two
different $G_0$ lines. In the later case, most terms are zero because
the two lines generally carry different momenta, and
$\sum_{\vec{q}_1,\vec{q}_2}^{}\epsilon_{\vec{q}_1}
\epsilon_{\vec{q}_2} =0$. Of 6 such terms generated in Eq. (\ref{ex}),
only one, coming from the outside $G_0$ lines of the non-crossed
diagram, is finite. The error from such terms decreases as one goes to
higher order diagrams, because the percentage of diagrams with one or
more pairs of $G_0$ lines of equal momenta decreases exponentially.
Similar arguments apply for higher powers in $t$. It follows that MA
captures most of the $t$ dependence of each diagram, while summing
over all diagrams. This suggests that MA may be accurate even far from
the limit $t\ll g$. Indeed, Eq. (\ref{sigma}) clearly shows that MA is
also valid for $g\ll t$.

For a better idea of the accuracy and range of the MA, we consider the
sum rules for $A(\vec{k},\omega)\equiv-{1\over \pi} \mbox{Im}
G(\vec{k},\omega)$, $M_n(\vec{k}) \equiv \int_{-\infty}^{\infty}d\omega
\omega^n A(\vec{k},\omega)$, which can be evaluated analytically. The usual
approach~\cite{Korni} is based on the equations of motion; since MA is
also based on them, it should fare well. Another approach~\cite{next}
is to start with the Dyson equation $G(\vec{k},\omega) =
G_0(\vec{k},\omega) + G_0^2(\vec{k},\omega) \Sigma(\vec{k},\omega) +
G_0^3(\vec{k},\omega) \Sigma^2(\vec{k},\omega) +\dots$ and the
perturbational expansion $\Sigma(\vec{k},\omega)=
g^2\Sigma^{(1)}(\vec{k},\omega) + g^4 \Sigma^{(2)}(\vec{k},\omega) +
\dots$, perform the integrals $ \int_{-\infty}^{\infty}d\omega
\omega^n G(\vec{k},\omega)$ and then take the imaginary part. This
task is aided by the fact that most terms in the integrand decay
faster than $1/\omega$ as $\omega\rightarrow \infty$, and their
contributions vanish. For $n=0,1$, only $G_0(\vec{k},\omega)$ has
finite contributions, giving $M_0(\vec{k})=1;
M_1(\vec{k})=\epsilon_{\vec{k}}$. In fact, $G_0$ contributes an
$\epsilon_{\vec{k}}^n$ to $M_n(\vec{k})$. Next is $G_0^2 g^2
\Sigma^{(1)}$. It decays like $1/\omega^3$, so it contributes only for
$n\ge 2$. Both SCBA and MA have the exact expression for
$\Sigma^{(1)}$, so they both satisfy exactly the $n=2$ and $3$ sum
rules. For $n\ge 4$, both $G_0^2 g^4 \Sigma^{(2)}$ and $G_0^3
\left(g^2 \Sigma^{(1)}\right)^2$ contribute. Since SCBA ignores one
$\Sigma^{(2)}$ diagram, it fails at this point, while MA is still
exact for $n=4$ and $5$. MA fails at $n=6$ because of the
approximations in diagrams' expressions. Instead of $M_6(\vec{k}) =
\epsilon_k^6+ g^2[5\epsilon_k^4+ 6t^4(2d^2-d)+4\epsilon_k^3\omega_E+
3\epsilon_k^2\omega_E^2+6dt^2(\epsilon_k^2+2\epsilon_k
\omega_E+2\omega_E^2)+2\epsilon_k\omega_E^3+\omega_E^4] +
g^4(18dt^2+12\epsilon_k^2 +22\epsilon_k\omega_E +25\omega_E^2)
+15g^6$, MA predicts a sum rule equal to $M_6(\vec{k})-2dt^2 g^4$ (the
dimension $d$ enters through $2dt^2 = {1\over N} \sum_{\vec{k}}^{}
\epsilon^2_{\vec{k}}$ and higher averages). In other words, MA
captures {\em exactly} both the dominant power in $t$, $\epsilon_k^n$, 
and the dominant power in $g$, which is $\sim g^n$ or $ \omega_Eg^{n-1}$, for
even/odd $n$. This also follows because MA is exact both for $t=0$
and $g=0$, so it can only miss terms $\sim g^4t^2$. These are lost
because terms from $G_0$ lines carrying equal momenta are neglected. As
discussed, such terms are a small minority of all contributions, and
indeed MA recovers the vast majority of terms in any $M_n$, like in
the $n=6$ case, showing that it is highly accurate not only for $t \ll
g$ and $t\gg g$, but also for intermediary values. By contrast,
although exact up to $n=3$, SCBA fails badly at higher $n$
because of the many higher order diagrams it neglects. For example, for
$n=6$, SCBA predicts $5g^6$ instead of $15g^6$ as the leading $g$ term
(with many ${\cal O}(g^4)$ terms missing), showing that SCBA fails for
$g\gg t$. Following this analysis we conclude that agreement with a few sum
rules is not meaningful; meaningful is to have agreement for the vast
majority of terms in {\em all sum rules}, and in particular for the
dominant terms in various limits. MA satisfies this restrictive condition.

MLF also captures both the $t=0$ and the $g=0$ limits
exactly. However, this alone does not suffice, either. Direct
evaluation shows that MLF fails badly all sum rules with $n\ge 1$ if
$g\ne 0$, as it predicts $M_1\rightarrow e^{-\lambda}\epsilon_k$, etc.

\begin{figure}[t]
\includegraphics[width=0.84\columnwidth]{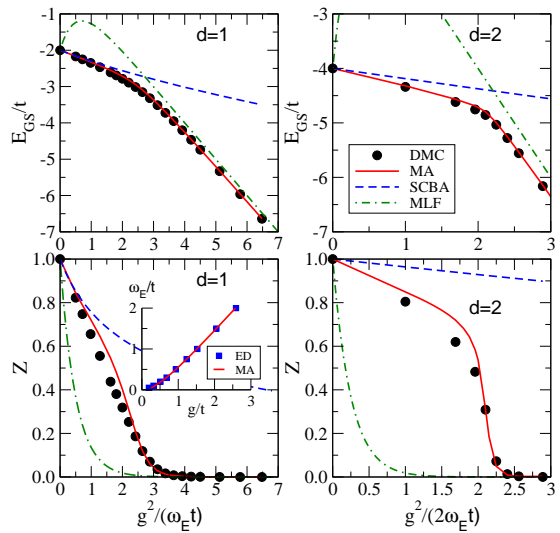}
\caption{GS energy, $E_{GS}$, and quasiparticle weight, $Z$,  for
  $\omega_E = 0.5 t$. $d=1$
  (left) and $d=2$ (right). Inset: line below
  which a second bound peak appears in 1d (see text). }
\label{fig2}
\end{figure}

A comparison of MA, SCBA and MLF against GS energies and qp weights
$Z= |\langle GS | c^{\dagger}_{k=0}|0\rangle|^2$ obtained with
DMC~\cite{Alex} are shown in Fig. \ref{fig2}, for both $d=1$ and 2. MA
compares equally well throughout the Brillouin zone~\cite{next}. The
agreement is best for $g\ll t$ and $g\gg t$, but is  good even for
$g\sim t$. In the inset, we show the line in parameter space below
which a second bound peak appears in 1d ({\em i.e.} the energy of the first
excited $k=0$ state satisfies $E_1 < E_{GS}+ \omega_E$). Above this
line, a continuum starts at $E_{GS}+ \omega_E$. The agreement between
MA and ED data of Ref.~\cite{EDvar} is excellent. By contrast, SCBA
never predicts a second bound state (it always finds the continuum);
MLF always predicts a peak at $E_{GS}+ \omega_E$, never a continuum.

\begin{figure}[t]
\includegraphics[width=0.91\columnwidth]{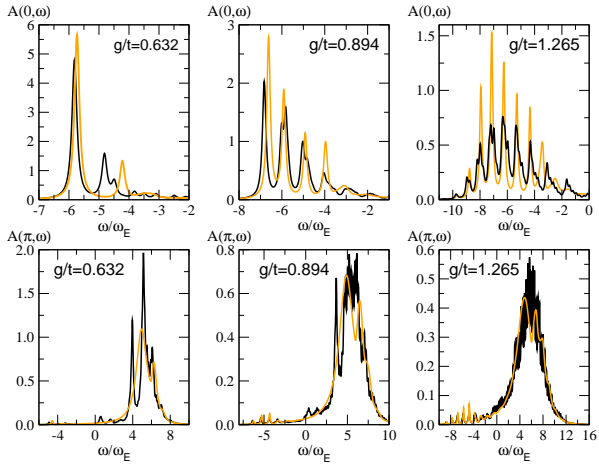}
\caption{(Color online) $A(k,\omega)$ in 1d, for $t=1,\omega_E=0.4,
\eta=0.1\omega_E$, $g=0.632, 0.894$ and 1.265, and $k=0$ and $\pi$.
MA data (orange line) vs. results from Ref. \cite{Cat} (black line). }
\label{fig3}
\end{figure}

A comparison of the MA spectral weight $A(k,\omega)$ (orange line) in 1d, for $k=0$
and $\pi/a$ and three values of $g$  is shown in
Fig. \ref{fig3}, against data (black line) based on a variational
method~\cite{Cat}. The agreement is again decent, especially since
$g\sim t$ in all three cases. Other comparisons with numerical data
are of similar quality~\cite{next}.

MA thus provides an accurate, yet simple and fast way to study
$G(\vec{k},\omega)$ for Holstein polarons in any dimension for
any $\epsilon_{\vec{k}}$.  The question is whether this approach can
be expanded beyond the Holstein Hamiltonian. While this issue is still
under investigation~\cite{next}, one possible route is provided by
this generalization of Eq. (\ref{12}):
\begin{equation}
\EqLabel{27} {\tilde \Sigma}(\vec{k},\omega) =
{1\over
  N}\!\!\sum_{\vec{q}_1,\lambda_1}\!\!\cfrac{E_1(\vec{k},\vec{q}_1,\lambda_1)}{\Large 1 
  - {1\over
	N}{\Large \sum_{\vec{q}_2,\lambda_2}}\cfrac{E_2(\vec{k},\vec{q}_1,\vec{q}_2, 
\lambda_1, \lambda_2)}{\large 1- \cdots}}   
\end{equation}
$E_1(\vec{k},\vec{q}_1,\lambda_1)\!=\!|g_{\vec{k},\vec{q}_1\lambda_1}|^2
G_0(\vec{k}_1,\omega_1)$, $E_2(\vec{k},\vec{q}_1,\vec{q}_2, \lambda_1,
\lambda_2)\!=\!|g_{\vec{k}_1,\vec{q}_2\lambda_2}|^2 G_0(\vec{k}_{1,2},
\omega_{1,2})\left( G_0(\vec{k}_1,\omega_1)+G_0(\vec{k}_2,
\omega_2)\right)$, etc, where $\vec{k}_i= \vec{k}-\vec{q}_i$,
$\vec{k}_{1,2}= \vec{k}-\vec{q}_1- \vec{q}_2$ and $\omega_i = \omega -
\omega_{\vec{q}_i,\lambda_i}$, $\omega_{1,2} = \omega -
\omega_{\vec{q}_1,\lambda_1} - \omega_{\vec{q}_2,\lambda_2}$.  This
describes coupling to several branches of bosons with dispersions
$\omega_{\vec{q}\lambda}$, with a vertex $g_{\vec{k},\vec{q}\lambda}$
if $\vec{k}$ is the momentum of the incoming particle and
$\vec{q}\lambda$ is the momentum and branch of the emitted
boson. While not exact, ${\tilde \Sigma}(\vec{k},\omega)$ is much more
accurate than $\Sigma_{MA}$: it gives all $1^{st}$, $2^{nd}$ and 9 out
of 10 of the $3^{rd}$ order diagrams exactly. The wrong $3^{rd}$
order diagram has one of its five $G_0$ lines with a wrong momentum
and energy. All higher order diagrams' numbers and topologies are
generated correctly, a small fraction of them having some mislabelled
$G_0$ lines. One can trace the first failing of a sum rule, due to the
wrong $3^{rd}$ order diagram, to now occur in $M_8(\vec{k})$.  For
simplicity, let us assume that boson frequencies $\omega_{\lambda}$
and vertices $g_{\lambda}$ depend only on the branch. Then, in
$M_8(\vec{k})$, a
term $ \langle|g_{\lambda}|^2\rangle
\langle\omega_{\lambda}|g_{\lambda}|^2\rangle^2$ is replaced by
$\langle|g_{\lambda}|^2\rangle^2\langle
\omega_{\lambda}^2|g_{\lambda}|^2\rangle$, and an extra term $2dt^2
\langle|g_\lambda|^2\rangle^3$ is generated, but all other terms
including the dominant $g$ terms $105 \langle|g_{\lambda}|^2\rangle^4$
are exact (here, $\langle f\rangle = \sum_{\lambda}^{}f$). Using a
further MA approximation removes the need to evaluate 
the momentum sums in (\ref{27}) by replacing all
$G_0(\vec{p},\Omega)\rightarrow \bar{g}_0(\Omega)$; this results in an
error in $M_6(\vec{k})$ (a missing
$2dt^2\langle|g_\lambda|^2\rangle^2$) but would speed up calculations
significantly and, as for the Holstein model, should have a limited
effect on accuracy. Further simplifications are possible if the boson
frequencies are close to each other. One can show that if $t=0$ and
all $\omega_\lambda=\omega_E$, Eq. (\ref{12}) with $g^2 =
\langle|g_\lambda|^2\rangle$ is the {\em exact}
self-energy~\cite{next}. We do not know if this identity has been
noted before. For close-by phonon energies, one can then also remove
the branch sums in (\ref{27}) by using Eq. (\ref{12}) with $g^2
=\langle|g_\lambda|^2\rangle$, $\omega_E = \langle\omega_\lambda
|g_\lambda|^2\rangle/g^2$, in which case one gets an error starting
with $M_4(\vec{k})$, where $\langle\omega_\lambda^2
|g_\lambda|^2\rangle\rightarrow g^2\omega_E^2$, but all dominant terms
are correct in all orders. Eq. (\ref{27}) may also be easy to estimate
for highly anisotropic $g_{\vec{q}}$, in which case the BZ
sums reduce to summations over a few hot spots. Of course, tests against numerical
results are needed to verify all this. 

To conclude, progress has been made in a very old problem,
by finding a simple yet highly accurate approximation for
$G(\vec{k},\omega)$ of the Holstein polaron. A path to possibly more
exciting results has also been uncovered.

{\bf Acknowledgments:} I thank George Sawatzky for suggesting this
problem and for many useful discussions, and A. Macridin,
V. Cataudella and G. De Filippis for sharing their numerical
results. This work was supported by NSERC and CIAR Nanoelectronics of
Canada.


\begin{thebibliography}{99}

\bibitem{Andrea} A. Damascelli, Z. Hussain, and Z.-X. Shen,
 Rev. Mod. Phys. {\bf 75}, 473 (2003).

\bibitem{hightc} K. M. Shen {\em et al.}, Phys. Rev. Lett. {\bf 93},
 267002 (2004).

\bibitem{Mahan} G. D. Mahan, {\em Many-Particle Physics}, (Plenum, New
  York, 1981).

\bibitem{QMC} N. V. Prokof'ev and B. V. Svistunov,
  Phys. Rev. Lett. {\bf 81}, 2514 (1998).

\bibitem{var} E. Jeckelmann and S. R. White, Phys. Rev. B {\bf 57},
  6376 (1998); A. H.  Romero, D. W. Brown, and K. Lindenberg,
  Phys. Rev. B {\bf 59}, 13728 (1999); V. Cataudella, G. De Filippis,
  and G. Iadonisi, Phys. Rev.  B {\bf 62 }, 1496 (2000).

\bibitem{EDvar} J. Bonca, S. A. Trugman, and I. Batistic, Phys. Rev. B
  {\bf 60}, 1633 (1999).

\bibitem{DMFT} S. Ciuchi {\em et al.}, Phys. Rev. B {\bf 56}, 4494
(1997).

\bibitem{Cat} G. De Filippis {\em et al.}, Phys. Rev. B {\bf 72},
014307 (2005).

\bibitem{Alexandr} A. S. Alexandrov and J. Ranninger, Phys. Rev. B
  {\bf 45}, 13109 (1992).

\bibitem{next} M. Berciu, unpublished.

\bibitem{Korni} P. E. Kornilovitch, EuroPhys Lett. {\bf 59}, 735
(2002).

\bibitem{Alex} A. Macridin, Ph.D. Thesis, Rijkuniversiteit
  Gr\"oningen, 2003.

\end{thebibliography}
\end{document}